\documentclass[prl,aps,twocolumn,superscriptaddress,longbibliography]{revtex4-2}

\usepackage[utf8x]{inputenc}
\usepackage[T1]{fontenc}
\usepackage{amssymb}
\usepackage{amsmath}
\usepackage{graphicx}
\usepackage{bbm}
\usepackage{psfrag}

\usepackage{latexsym}
\usepackage{color}
\usepackage[dvipsnames]{xcolor}
\usepackage{hyperref}
\usepackage{orcidlink}
\hypersetup{colorlinks=true, citecolor=blue, urlcolor=blue, linkcolor=blue,urlcolor=cyan}
\usepackage{tensor}
\usepackage{amsfonts}
\allowdisplaybreaks[4]        
\usepackage{euscript}           
\usepackage{tensor}        
\usepackage{amsthm} 
\usepackage{subfigure}
\usepackage{bbm}
\usepackage[header,title,page,titletoc]{appendix}  
\usepackage[most]{tcolorbox}
\tcbuselibrary{skins, breakable, theorems}
\usepackage{verbatim}

\usepackage{bm}

\usepackage[final]{microtype}
\microtypesetup{protrusion=true, expansion=true}

\newcommand{\dd}{\mathrm{d}}

\renewcommand{\[}{\left[}

\def\GeV{\,\mathrm{GeV}}
\def\pc{\,\mathrm{pc}}
\def\kpc{\,\mathrm{kpc}}
\def\Gyr{\,\mathrm{Gyr}}

\begin{document}

\author{Chen Zhang\,\orcidlink{0000-0003-2649-8508}}\email[Contact author: ]{zhangchen2@mail.neu.edu.cn}
\author{Chun-Yan Jiang}
\author{Nayun Jia}
\affiliation{\it Liaoning Key Laboratory of Cosmology and Astrophysics, College of Sciences,
	Northeastern University, No.~3-11, Wenhua Road, Heping District, Shenyang 110819, China}
\author{Xin Zhang\,\orcidlink{0000-0002-6029-1933}}\email[Contact author: ]{zhangxin@neu.edu.cn}
\affiliation{\it Liaoning Key Laboratory of Cosmology and Astrophysics, College of Sciences,
	Northeastern University, No.~3-11, Wenhua Road, Heping District, Shenyang 110819, China}
\affiliation{\it MOE Key Laboratory of Data Analytics and Optimization for Smart Industry,
	Northeastern University, No.~3-11, Wenhua Road, Heping District, Shenyang 110819, China}
\affiliation{\it National Frontiers Science Center for Industrial Intelligence and Systems Optimization,
	Northeastern University, No.~3-11, Wenhua Road, Heping District, Shenyang 110819, China}

\title{Self-Consistent Parker Bound on Magnetic Monopoles}

\begin{abstract}
Magnetic monopoles arise generically in unified theories and offer a natural explanation of charge quantization. Beyond collider searches and cosmic-ray experiments, their flux is constrained by Parker-type bounds requiring galactic magnetic fields to survive monopole energy extraction. We formulate a self-consistent Parker bound anchored in the lowest eigenmode of the galactic mean-field dynamo and convert the resulting limit to the present-day flux. Small-scale turbulent fields both seed this eigenmode and set the monopole velocity via stochastic acceleration before energy extraction from the coherent field. These unavoidable effects substantially modify the standard extended Parker bound at low and intermediate masses, yielding flux limits robust to primordial magnetic fields (PMFs); PMFs strong enough to alter these limits lie in regimes constrained by Ly$\alpha$ data or testable by 21-cm observations and other cosmological probes.
\end{abstract}

\maketitle

\noindent \emph{1.~\textbf{Introduction}.—}
Magnetic monopoles are hypothetical particles that would symmetrize Maxwell's equations under electric-magnetic duality~\cite{Preskill:1984gd,Rajantie:2012xh,Patrizii:2015uea,Mavromatos:2020gwk,Mitsou:2026zcf,Vilenkin:2000jqa,Shnir:2005vvi,Weinberg:2012pjx}. Dirac showed that consistency with quantum mechanics requires the magnetic charge $g$ to be an integer multiple of the Dirac charge $g_D\simeq68.5e$, thereby explaining charge quantization~\cite{Dirac:1931kp}. In grand unified theories (GUTs) and partially unified scenarios such as Pati-Salam and trinification models~\cite{Georgi:1974sy,Langacker:1980kd,Lazarides:1980cc,Lazarides:2019xai,Pati:1974yy,Hartmann:2014fya,DiLuzio:2020xgc,Dolan:2020doe,Kephart:2025tik,Senjanovic:2025enc,Kephart:2006zd,Kephart:2017esj,Lazarides:2021tua,Brennan:2024sth,Lazarides:2024niy,Maji:2026nkz}, monopoles generically arise as topological defects~\cite{tHooft:1974kcl,Polyakov:1974ek}, with masses tied to the unification scales. Cosmological phase transitions~\cite{Mazumdar:2018dfl,Hindmarsh:2020hop,Athron:2023xlk} can produce them through the Kibble or Kibble-Zurek mechanism~\cite{Kibble:1976sj,Zurek:1985qw,Murayama:2009nj,delCampo:2013nla,Graesser:2020hiv,Patel:2021iik,Baek:2013dwa,Khoze:2014woa,Kawasaki:2015lpf,Nomura:2015xil,Bai:2020ttp,Fan:2021ntg,Yang:2022quy}, with their surviving abundance determined by phase transition dynamics and defect evolution~\cite{Zeldovich:1978wj,Preskill:1979zi,Guth:1979bh,Goldman:1980sn,Dicus:1982ri,Weinberg:1983uq,Martins:2008ks,Sousa:2017wvx,Zhang:2023tfv,Zhang:2023zmb,Hindmarsh:2025vxh,Dvali:1997sa,Stojkovic:2004hz,Brush:2015vda,Dvali:2022vwh,Bachmaier:2023zmq,Bachmaier:2025jaz}.
Although superheavy GUT monopoles are usually diluted by inflation in viable cosmologies, lighter monopoles with masses $m\lesssim 10^{17}\GeV$ can be produced after inflation in well-motivated partially unified theories, leaving fluxes accessible to direct searches in cosmic rays~\cite{MACRO:2002jdv,PierreAuger:2016imq,IceCube:2021eye,TelescopeArray:2023sbd,Cho:2023krz,Frampton:2024shp,SCEP:2024cir,Gould:2024zed,Candela:2025gwp,Perri:2025qpg}. Sufficiently light monopoles can be produced and probed in high-energy $pp$ and heavy-ion collisions~\cite{Gould:2017zwi,ATLAS:2023esy,MoEDAL:2021vix,MoEDAL:2024wbc}, and also in cosmic-ray interactions with the atmosphere or interstellar medium~\cite{Iguro:2021xsu,Iguro:2024oml}. Astrophysical systems also provide powerful probes of magnetic charges. Examples include searches for a monopole component of Earth’s magnetic field~\cite{Bai:2021ewf}, and constraints on magnetic black holes and milli-magnetic monopoles from white dwarfs or magnetars~\cite{Hook:2017vyc,Maldacena:2020skw,Bai:2020ezy,Pereniguez:2024fkn,Graesser:2021vkr,Bai:2020spd,Ghosh:2020tdu,Diamond:2021scl}.

A classic example of an astrophysical probe is the Parker bound, based on the survival of a coherent galactic magnetic field in the presence of a flux of magnetic monopoles~\cite{Parker:1970xv,Turner:1982ag,Adams:1993fj,Kobayashi:2023ryr,Zhang:2024mze}. An excessive flux would dissipate the field faster than it can be regenerated. Conventional implementations, however, usually treat the field strength, coherence length, regeneration time, and monopole velocity as effective inputs, leaving substantial freedom in the very quantities that define the bound. This freedom reflects not only astrophysical uncertainty, but also the lack of a well-defined formulation of the bound within modern theories of galactic magnetism. These theories describe a staged evolution involving turbulent amplification, seed-field formation, mean-field eigenmode growth, and large-scale ordering~\cite{Ruzmaikin:1988mfg,Shukurov:2022amf,Brandenburg:2004jv,Beck:2015jta,Brandenburg:2023mvu,Arshakian:2008cx,Arshakian_2011,Rodrigues_2019,Moss1998BoundaryEffects}, in which the quantities entering the Parker argument are linked rather than independent. The task is therefore not merely to refine the Parker estimate, but to close it by specifying the magnetic structure and evolutionary stage that define the bound.

\begin{figure}[t]
\centering\includegraphics[width=3.37in]{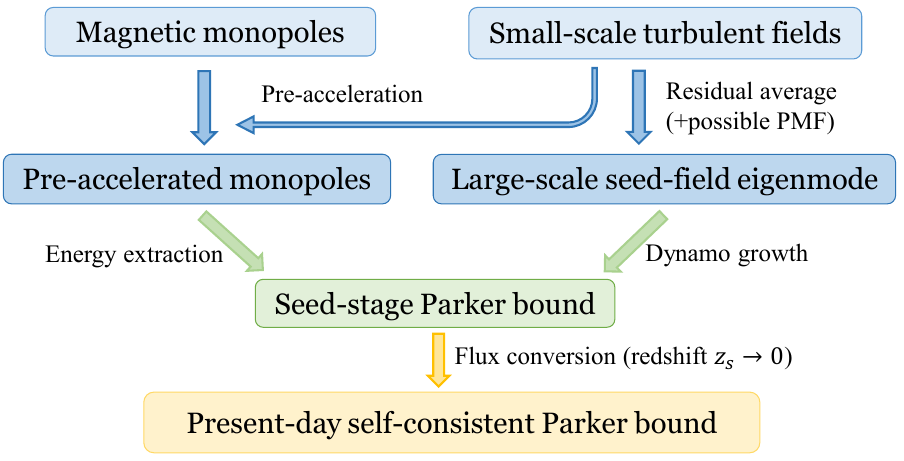}
\caption{Schematic flow of the self-consistent Parker-bound calculation, including a possible PMF contribution to the seed-field eigenmode.}\label{fig:FlowDiagram}
\end{figure}

In this Letter, we formulate such a self-consistent Parker bound. The magnetic reservoir is identified with the lowest seed eigenmode of the galactic mean-field dynamo, whose spatial localization and growth rate define the field-survival problem. Turbulent magnetic fields whose residual supplies the seed also accelerate monopoles, fixing the velocity entering the energy extraction rather than treating it as an external input. The seed-stage flux bound is then converted to the present-day flux by following subsequent monopole evolution. The resulting calculation flow is summarized in Fig. ~\ref{fig:FlowDiagram}. Primordial magnetic fields (PMFs), if present, can be incorporated as an additional projection onto the seed eigenmode. The construction ties field strength, coherence length, regeneration time, monopole velocity, and flux conversion to one evolutionary history, yielding a closed consistency condition for the survival of galactic seed fields and substantially altering the Parker limits at low and intermediate masses.

\begin{figure}[t]
	\centering\includegraphics[width=3.37in]{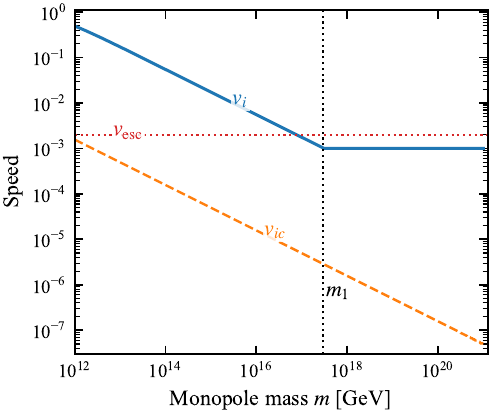}
	\caption{Monopole speed $v_i$ after turbulent pre-acceleration (blue solid). The orange dashed and red dotted curves denote the	critical and escape speeds, $v_{ic}$ and $v_{\rm esc}=2v_{i0}$; $m_1$ marks equal turbulent and initial kinetic energies.}\label{fig:PreVelocity}
\end{figure}

\noindent \emph{2.~\textbf{Turbulent pre-acceleration of monopoles}.—}
Small-scale turbulent magnetic fields are a generic component of galactic magnetism. They are the random magnetic component amplified and maintained primarily by the fluctuation dynamo, which stretches and folds an initially weak field through turbulent motions driven by gravitational collapse and virialization in protogalactic halos and by supernova feedback, gas accretion, mergers, and other stirring processes in later galaxies. This random component is expected to remain important throughout a galaxy's life~\cite{Brandenburg:2023mvu} and, through its residual average over finite galactic volumes, to seed the subsequent large-scale mean-field dynamo~\cite{Ruzmaikin:1988mfg}. At this seeding stage, it can be modeled as filling a galactic disk of radius $r_0$ with many turbulent magnetic cells of characteristic field strength $b_0$ and coherence length $l_{\rm eff}$, which stochastically accelerate monopoles. A monopole of mass $m$ and magnetic charge $g$, entering the turbulent region with Lorentz factor $\gamma_{i0}\equiv(1-v_{i0}^2)^{-1/2}$~\footnote{In this work we adopt the natural units, so that the speed of light in vacuum is set to $c=1$.}, emerges with a typical Lorentz factor $\gamma_i\equiv(1-v_i^2)^{-1/2}$ determined by~\cite{Kobayashi:2023ryr}
\begin{align}
m(\gamma_i-1)\simeq \max\{m(\gamma_{i0}-1),\,(N_t/2)^{1/2}g b_0 l_{\rm eff}\},\label{eqn:preacceleration}
\end{align}
if it traverses $N_t=r_0/l_{\rm eff}$ cells. In \hyperref[sec:appA]{Appendix A} of the Supplemental Material it is shown that $l_{\rm eff}=4L_L=(3/2)L_{\rm int}$ with
$L_L$ and $L_{\rm int}$ being two distinct magnetic integral scales defined in ~\cite{Shukurov:2022amf}. Simulations of fluctuation dynamos suggest $L_{\rm int}\simeq (1/4\text{--}1/3)L_u$ at saturation with $L_u\simeq 100\pc$ being the velocity integral scale~\cite{Shukurov:2022amf,Bhat:2012sb}, which we identify with the outer scale of the turbulence. As a result, we take $l_{\rm eff}=50\pc$ and for numerical purposes use the fiducial set of parameters $g=g_D,v_{i0}=10^{-3}$ and $b_0=5\,\mu\mathrm{G},r_0=9\kpc$ with which we obtain $v_i$ as a function of $m$ shown in Fig.~\ref{fig:PreVelocity}~\footnote{We restrict ourselves to non-relativistic monopoles and thus do not consider the mass range $m<10^{12}\GeV$.}. This turbulent pre-acceleration significantly increases the incident speed above $v_{i0}$ for $m\lesssim 3\times10^{17}\,{\rm GeV}$, thereby reducing the deviation of monopoles by the large-scale field. As we will see, the amount of this deviation has a crucial impact on the flux upper limits. Therefore turbulent pre-acceleration is an essential part of the self-consistent closure of the bound.

\noindent \emph{3.~\textbf{Energy extraction from the seed eigenmode}.—}
Conventional implementations of the Parker bound rely on two crucial parameters of the large-scale galactic magnetic field: the coherence length and the regeneration time. However, in practice their operational definitions are somewhat vague. The effective coherence length is actually a manifestation of front propagation, resulting from interaction among multiple dynamo eigenmodes~\cite{Moss1998BoundaryEffects}. The regeneration time can be related to the inverse growth rate of some dynamo eigenmode, but sometimes the longer ordering timescale of the coherent field is used instead~\cite{Bai:2020spd,Diamond:2021scl,Zhang:2024mze}. For the purpose of setting up the Parker bound, we propose a self-consistent treatment in which well-defined field amplification parameters are to be used instead of ordering parameters. This means the Parker bound should be defined with respect to the lowest eigenmode of the galactic mean-field dynamo. Its width of localization can be taken as the coherence length and its inverse growth rate should be taken as the field regeneration time. Since the eigenmode must survive already from its seed stage before subsequent amplification and ordering can occur, we consider the seed stage of this eigenmode, with the seed-field amplitude $\tilde{B}$ determined by the residual average of turbulent magnetic fields in its localization region~\cite{Rodrigues_2019,Ruzmaikin:1988mfg}:
\begin{align}
\tilde{B} \simeq \frac{1}{3}\frac{b_0}{N_l^{1/2}}\frac{L_{\rm int}}{\Delta r},
\quad
N_l=\frac{3h_0 r_s \Delta r}{L_{\rm int}^3}. \label{eqn:SeedField}
\end{align}
Here the localization region of the lowest eigenmode can be viewed as an annular cylindrical shell of thickness $2h_0$ ($h_0$ is the disk half-thickness), localization radius $r_s$ and width $\Delta r\simeq\sqrt{r_0 h_0}$~\footnote{A derivation of the relation $\Delta r\simeq\sqrt{r_0 h_0}$ is given in \hyperref[sec:appB]{Appendix B} of the Supplemental Material.}, and $N_l$ denotes the number of turbulent magnetic field cells (of size $L_{\rm int}$) contained in the localization region. The factor $\frac{L_{\rm int}}{\Delta r}$ accounts for the solenoidality condition for turbulent magnetic fields~\cite{Rodrigues_2019,Ruzmaikin:1988mfg}. Although other contributions to the seed field might exist, such as PMFs, the above contribution from turbulent magnetic fields is an irreducible one, and is expected to overwhelm contributions directly from usual battery mechanisms. The same turbulent fields determine monopole pre-acceleration and the seed-field amplitude for the mean-field dynamo eigenmode, constituting one further important ingredient for the closure of the self-consistent Parker bound. Using $h_0=500\pc,r_s=6\kpc$ in addition to the previous fiducial set of parameters, we obtain $\tilde{B} \simeq 3.64\times 10^{-11}\,\mathrm{G}$.

The active turbulent magnetic fields do not interfere with the Parker bound argument for the mean-field dynamo eigenmode, since the interference contribution is mostly cancelled when a monopole traverses many uncorrelated cells, and averages to zero when a large number of monopoles are considered~\footnote{This is demonstrated in \hyperref[sec:appC]{Appendix C} of the Supplemental Material.}. Then, the energy of the mean-field dynamo eigenmode with field strength $B$ will be extracted by an amount $\Delta E=\frac{1}{4}\frac{(gB\,\Delta r)^2}{m(\gamma_i-1)}$ in the slight deviation case or $\Delta E=gB\Delta r$ in the significant deviation case~\cite{Kobayashi:2023ryr}. By slight or significant deviation we mean whether a monopole changes its velocity slightly or significantly during its journey in the localization region after pre-acceleration. The boundary dividing slight and significant deviation cases is determined via $4m(\gamma_i-1)=gB\Delta r$ and is shown for the monopole critical velocity $v_{ic}$ based on the seed field $\tilde{B}$ in Fig.~\ref{fig:PreVelocity} as the orange dashed line. It is clear that $v_{ic}<v_i$ by orders of magnitude, implying that energy extraction from the seed field by monopoles is always in the slight deviation regime due to the unavoidable turbulent pre-acceleration.

\noindent \emph{4.~\textbf{Self-consistent Parker bound}.—}
The evolution of the mean-field dynamo eigenmode strength $B$ can be modeled by (in Gaussian units) 
\begin{align}
\frac{\mathrm{d}B}{\mathrm{d}t}
=\gamma B-\alpha B^2
-12\pi^2 Q^{-1}\,\frac{F_s g}{1+2\mu/B},\label{eqn:Bevolve}
\end{align}
where $\mu\equiv\frac{2m(\gamma_i-1)}{g\,\Delta r}$ and $F_s$ denotes the monopole flux at the seed stage. The $\gamma B$ term embodies the mean-field dynamo amplification with $\gamma^{-1}=\tau_B$ being the field regeneration time of the lowest eigenmode. Note that $\tau_B$ should be related to the energy regeneration time $\tau_E$ implicitly assumed in previous studies by $\tau_B=2\tau_E$. The $-\alpha B^2$ term is introduced to allow the field strength to saturate at the observed value~\cite{Adams:1993fj}. The last term in Eq.~\eqref{eqn:Bevolve} characterizes energy extraction by monopoles, which can be derived using the expressions for $\Delta E$ obtained previously in the slight and significant deviation cases. $Q$ is a geometric factor related to the shape of the localization region of the eigenmode, which is found to be $Q=\frac{3V}{A\Delta r}$ with $A$ and $V$ being the surface area and the volume of the annular cylindrical shell under consideration~\footnote{The validity of considering a nonspherical magnetized region for the Parker bound, including the derivation of the relevant geometric factor, is shown in \hyperref[sec:appD]{Appendix D} of the Supplemental Material.}. We have approximately $Q=\frac{3h_0}{\Delta r+2h_0}$. If the eigenmode starts with the seed-field strength $\tilde{B}$, then using the same analytic argument for Eq.~\eqref{eqn:Bevolve} as in ~\cite{Adams:1993fj}, we conclude that the survival of $B$ is maintained if and only if $F_s<F_s^{\rm up}$ with
\begin{align}
F_s^{\rm up} = \frac{Q}{12\pi^2 g} \left[ \frac{4m(\gamma_i-1)}{g\Delta r} +\tilde{B} \right]\tau_B^{-1}. \label{eqn:Fsup}
\end{align}
A tiny dependence of $F_s^{\rm up}$ on $\alpha$ has been neglected, which is legitimate since $\tilde{B}$ is much smaller than the saturation value of $B$. Note that the Lorentz factor that appears in Eq.~\eqref{eqn:Fsup} is $\gamma_i$, which encodes the effect of turbulent pre-acceleration. 
Replacing the $\max$ in Eq.~\eqref{eqn:preacceleration} by a smooth conservative sum, the seed-stage flux upper limits can be expressed as
\begin{align}
F_s^{\rm up}
&= \frac{1}{12\pi^2}\frac{3h_0}{\Delta r+2h_0}\,
g^{-1}\left[\frac{4m(\gamma_{i0}-1)}{g\Delta r}\right. \notag\\[-0.2ex]
&\quad \left.
+2\sqrt{3}\left(\frac{\kappa L_u}{h_0}\right)^{1/2} b_0
+\tilde B\right]\tau_B^{-1}. \label{eqn:Fsup2}
\end{align}
The second term in the square brackets is due to turbulent pre-acceleration, and is now expressed in terms of the velocity integral scale $L_u$ and $\kappa\equiv\frac{L_{\rm int}}{L_u}\simeq\frac{1}{3}$. In obtaining this form we have again used $N_t=r_0/l_{\rm eff}$, $l_{\rm eff}=(3/2)L_{\rm int}=(3/2)\kappa L_u$ and $\Delta r\simeq\sqrt{r_0 h_0}$.

In order to facilitate comparison with the direct search limits, the seed-stage flux upper limits $F_s^{\rm up}$ should be converted to the present-day flux upper limits $F^{\rm up}\equiv\mathcal{C}F_s^{\rm up}$, with $\mathcal{C}$ being the conversion factor. The conversion is based on $F=\frac{nv}{4\pi}$ with $n$ and $v$ being the monopole number density and velocity incident on the eigenmode localization region, respectively. Since monopoles have a small charge-to-mass ratio, we assume the monopole distribution traces that of dark matter if astrophysical magnetic fields are negligible. The main factors contributing to the flux conversion are then due to acceleration by turbulent and large-scale galactic magnetic fields. Let $v_i$ be the monopole velocity after pre-acceleration by turbulent magnetic fields, $v_L$ be the monopole velocity after acceleration by large-scale coherent fields, and $v_\mathrm{esc}$ be the escape velocity of the galaxy. Since the turbulent and large-scale coherent fields have comparable saturation strengths, we expect $v_i<v_L$ to always hold. If $v_L<v_\mathrm{esc}$, then monopoles always remain bound to the galaxy despite the acceleration by turbulent and large-scale magnetic fields, thus $n$ and $v$ are not expected to vary significantly and we take $\mathcal{C}\simeq 1$. If $v_i>v_\mathrm{esc}$, then monopoles initially bound to the galaxy are expected to escape due to turbulent pre-acceleration alone. In such a case, both the monopole flux that the seed-stage Parker bound constrains and the present-day monopole flux receive their main contributions from the diffuse component of the Local Group. $n$ and $v$ of the diffuse component are not expected to vary significantly, and thus we again estimate $\mathcal{C}\simeq 1$. The last but most interesting possibility is if $v_i<v_\mathrm{esc}<v_L$, then monopoles initially bound to the galaxy cannot escape by turbulent pre-acceleration alone, but can escape due to acceleration in large-scale coherent fields. Then the seed-stage monopole flux is mainly from bound monopoles, while the present-day monopole flux is mainly from the diffuse component of the Local Group. In this case $\mathcal{C}\simeq\frac{\Delta_\mathrm{LG}}{\Delta_\mathrm{G}}$, with $\Delta_\mathrm{LG}$ and $\Delta_\mathrm{G}$ denoting the dark matter overdensity of the Local Group and the eigenmode localization region in the galaxy, respectively. In this last case, the fiducial value of $\mathcal{C}$ is set to be $10^{-2}$, and relaxed to $10^{-1}$ if a conservative estimate is needed~\cite{Baushev:2012dm,Kakharov:2025myy}.
\begin{figure}[t]
	\centering\includegraphics[width=3.37in]{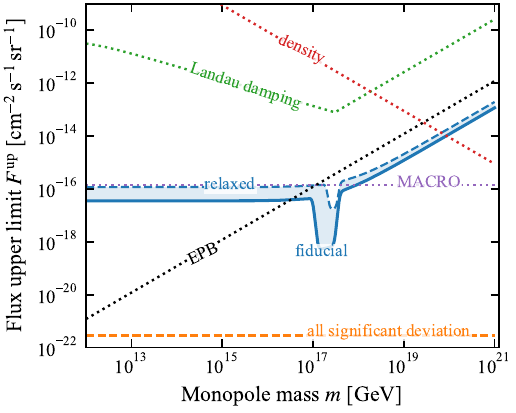}
	\caption{Present-day self-consistent Parker flux limits for the	fiducial (blue solid) and relaxed (blue dashed) parameter sets.	Also shown are the result assuming significant deviation at all masses, the MACRO direct-search limit~\cite{MACRO:2002jdv}, the density bound for clustered monopoles, the conventional extended Parker bound (EPB)~\cite{Adams:1993fj}, and the boundary of strong Landau damping.}\label{fig:FluxLimits}
\end{figure}

The present-day self-consistent Parker bound is shown in Fig.~\ref{fig:FluxLimits} as a blue solid curve using the fiducial set of input parameters for the Milky Way:
\begin{align}
\tau_B & =0.5\Gyr,b_0=5\,\mu\mathrm{G},h_0=500\pc,L_u=100\pc,\nonumber \\
r_0 & =9\kpc,\kappa=1/3,r_s=6\kpc,v_{i0}=10^{-3},g=g_D, \label{eqn:fiducial}
\end{align}
and the conversion factor $\mathcal{C}$ is set to $10^{-2}$ in the $v_i<v_\mathrm{esc}<v_L$ case and $1$ otherwise (we use $v_\mathrm{esc}=2v_{i0}$), with the junction region smoothed with a hyperbolic tangent function. Except for the conversion factor, the self-consistent Parker bound exhibits a two-regime mass dependence similar to conventional implementations of the Parker bound. Nevertheless, the reason behind the flattening behavior in the low-mass range is completely different. For the self-consistent Parker bound, the flattening is not because the monopole acceleration enters the significant deviation regime, but rather due to turbulent pre-acceleration by small-scale magnetic fields which contribute much more than $\tilde{B}$ in Eq.~\eqref{eqn:Fsup2}. This leads to significant relaxation of the standard extended Parker bound (black dotted line in Fig.~\ref{fig:FluxLimits}) in the low-mass range~\footnote{Consequently, the current astrophysical constraint in this mass range is only slightly stronger than or comparable to the MACRO direct-search limit. Absent substantial improvement in the astrophysical treatment, future large-area nuclear track detector searches, such as the array proposed in Ref.~\cite{Gould:2024zed}, could therefore set the most stringent flux limits in this region of parameter space.}. In the high-mass range, the self-consistent Parker bound is dominated by monopole kinetic energy before turbulent pre-acceleration, and is similar to the standard extended Parker bound with slight improvements due to refined treatments in certain $\mathcal{O}(1)$ factors and parameters. In Fig.~\ref{fig:FluxLimits} the blue dashed curve presents a more conservative bound obtained by setting $\tau_B=0.3\Gyr,b_0=10\,\mu\mathrm{G}$ and $\mathcal{C}=10^{-1}$ in the $v_i<v_\mathrm{esc}<v_L$ case, with other parameters unchanged. This reasonable variation of astrophysical inputs provides an estimate of uncertainties involved in interpreting the results. In most of the mass range the bound is fairly robust against the variation in parameters, except for a small range near $10^{17}\GeV$ where the uncertainty associated with flux conversion is large. The validity of the Parker bound requires monopole plasma oscillations to exhibit strong Landau damping which occurs when $F\lesssim\frac{1}{4}mv_i^3g^{-2}(\Delta r)^{-2}$~\cite{Zhang:2024mze}. The boundary of strong Landau damping is shown in Fig.~\ref{fig:FluxLimits} by the green dotted curve, safely above the self-consistent Parker bound. A monopole flux above the boundary should be rejected based on arguments presented in ~\cite{Parker:1987monopole_plasma}.

\noindent \emph{5.~\textbf{Robustness to primordial magnetic fields}.—} PMFs can be produced in the early universe via inflationary magnetogenesis or phase transitions~\cite{Widrow:2002ud,Subramanian:2015lua,Subramanian:2019jyd,Vachaspati:2020blt}, long before galactic magnetism starts to operate. Blazar observations set a lower limit $\sim 10^{-16}\mathrm{G}$ on the strength of intergalactic magnetic fields~\cite{Neronov:2010gir,Essey:2010nd,Long:2015cza,Finke:2015ona,Fermi-LAT:2018jdy} which could be a manifestation of PMFs. The Parker bound can be affected by PMFs in two ways. The first is through an even earlier pre-acceleration of monopoles in intergalactic voids by PMFs. From ~\cite{Perri:2023ncd}, we infer that for the range of monopole mass under consideration and the flux level to be constrained in this work, the effect of PMF pre-acceleration is negligible due to backreaction of monopoles on PMFs. The second is through the impact of PMFs on the initial conditions of galactic magnetism. PMFs are amplified and stirred during the collapse of the protogalaxy. Their impact can be analyzed via further projection onto eigenmodes of small-scale and large-scale dynamos. On small scales, the projection may modify the seed field amplitude for the fluctuation dynamo. Since the fluctuation dynamo gets saturated in a short period of time, PMFs can hardly affect the magnetic evolution on small scales. On large scales associated with the mean-field dynamo, the projection may again modify the seed field amplitude. This may alter the flux upper limits in principle, by adding a new contribution $\tilde{B}_\star$ to the square brackets of Eq.~\eqref{eqn:Fsup2}. The self-consistent Parker bound would not be affected for $\tilde{B}_\star\lesssim 2\sqrt{3}(\frac{\kappa L_u}{h_0})^{1/2} b_0$, that is $\tilde{B}_\star\lesssim 4.47\,\mu\mathrm{G}$ based on Eq.~\eqref{eqn:fiducial}. If PMFs have a nearly scale-invariant spectrum $P_B(k)\propto k^{n_B},n_B=-3+\epsilon$ with a small positive $\epsilon$, such as in inflationary magnetogenesis, this condition can be translated into the following requirement on the comoving strength of the PMF smoothed on the $1\,\mathrm{Mpc}$ scale:
\begin{align}
B_{1\mathrm{Mpc}} < \frac{4.47\times10^{-6}\,\mathrm{G}}
{\chi C_{\mathrm{coll}}}
\left(\frac{\lambda_P}{1\,\mathrm{Mpc}}\right)^{\epsilon/2}, \label{eqn:PMF1}
\end{align}
in which $\chi<1$ is a geometric factor accounting for the projection of the processed PMFs onto the lowest eigenmode of the mean-field dynamo, while $C_{\mathrm{coll}}$ and $\lambda_P$ can be expressed using the gas number density $n_g$ at the seed stage and baryon number density today $\bar{n}_{b0}$ as
\begin{align}
C_{\mathrm{coll}}=\left(\frac{n_g}{\bar{n}_{b0}}\right)^{2/3},\quad\lambda_P=\left(\frac{n_g}{\bar{n}_{b0}}\right)^{1/3}\Delta r.
\end{align}
$C_{\mathrm{coll}}\simeq 10^3\text{--}10^4$ accounts for the flux freezing during the protogalaxy collapse with the redshift factor stripped off. $\lambda_P\simeq 100\kpc$ is the scale obtained by unfolding $\Delta r$ to the protogalaxy stage and then redshifting to today. Violation of Eq.~\eqref{eqn:PMF1} would require $B_{1\mathrm{Mpc}}\gtrsim\mathcal{O}(0.1)\,\mathrm{nG}$, a regime already constrained by Ly$\alpha$-forest data and accessible to 21-cm or other astrophysical probes~\cite{Cruz:2023rmo,Bhaumik:2024efz,Pavicevic:2025gqi,Bhaumik:2026mlc,Worku:2026sbp,Aramburo-Garcia:2022ywn}. For causal PMFs with a typical Batchelor spectrum $P_B(k)\propto k^2$, the counterpart of Eq.~\eqref{eqn:PMF1} constrains the comoving effective field strength $B_I$ at the integral scale $\lambda_I$, leading to the conclusion that causal PMFs would threaten the bound only if they were $\mathcal{O}(1\text{--}10)\,\mathrm{nG}$ on $\sim 10\text{--}100\kpc$ scales, or even stronger on smaller phase-transition scales, a parameter region already strongly constrained or soon testable by cosmological probes~\cite{Planck:2015zrl,Sutton:2017jgr,Paoletti:2022gsn,Cruz:2023rmo,Pavicevic:2025gqi}.

\noindent \emph{6.~ \textbf{Conclusion}.—}
We have formulated a Parker bound whose magnetic reservoir is the lowest seed eigenmode of the galactic mean-field dynamo. Its localization width and inverse growth time fix the coherence scale and regeneration time, while the turbulent field that seeds the eigenmode also stochastically pre-accelerates monopoles. The resulting energy extraction remains in the slight-deviation regime, so the low-mass behavior is governed by turbulent pre-acceleration rather than by conventional significant-deviation saturation. Converting the seed-stage constraint to the present epoch yields a self-consistent Parker bound that is substantially relaxed relative to the standard extended bound at low and intermediate masses, and approaches it at high masses, apart from controlled order-unity refinements. The result is robust to reasonable galactic uncertainties and to PMFs unless their processed projection rivals the contribution from turbulent pre-acceleration, a regime already constrained or testable by Ly$\alpha$-forest, 21-cm, and other cosmological probes. This closes the Parker argument by embedding it in the same dynamo history that generates the constrained field, and thus exposes a direct interface between monopole phenomenology, evolution of galactic magnetic fields, and future probes of primordial magnetism. It also provides a concrete framework for systematic refinements of the Parker bound through more realistic dynamo modeling, galaxy-formation simulations, and observations of magnetic fields across cosmic time.

\vspace{8pt}

\begin{acknowledgments}

\noindent \emph{Acknowledgments:}
We thank Rui-Ning Zhang for her participation in the early stage of this project. C. Zhang would like to thank Yue Shao for discussions on 21-cm probes of PMFs. This work is supported by the National Natural Science Foundation of China (Grants Nos.\ 12533001, 12575049, 12473001, and 12447105), the National SKA Program of China (Grants Nos.\ 2022SKA0110200 and 2022SKA0110203), the China Manned Space Program (Grant No.~CMS-CSST-2025-A02), and the 111 Project (Grant No.\ B16009).
C.~Zhang is supported by the Joint Fund of Natural Science Foundation of Liaoning Province (Grant No.\ 2023-MSBA-067) and the Fundamental Research Funds for the Central Universities (Grant No.\ N2405011).
\end{acknowledgments}

\bibliography{references}


\setcounter{equation}{0} 
\renewcommand{\theequation}{S\arabic{equation}} 

\titlepage

\setcounter{page}{1}

\begin{center}
{\LARGE Supplemental Material}
\end{center}

\section{Appendix A. Coherence Lengths of Turbulent Magnetic Fields}\label{sec:appA}

In the literature, two integral scales $L_\mathrm{int}$ and $L_L$ are introduced to characterize turbulent magnetic fields~\cite{Shukurov:2022amf}
\begin{align}
L_L = \frac{\int_0^\infty M_L(r)\,\mathrm{d}r}{M_L(0)},\,\,L_{\rm int}=2\pi\,\frac{\int_{0}^{\infty} k^{-1}M(k)\,\mathrm{d}k}{\int_{0}^{\infty} M(k)\,\mathrm{d}k},
\label{eqn:IntScales}
\end{align}
with $M_L(r)$ being the longitudinal magnetic correlation function and $M(k)$ being the one-dimensional magnetic energy spectrum. These two scales are related by
\begin{align}
L_{\rm int}=\frac{8}{3}L_L.
\end{align}

On the other hand, in the derivation of the self-consistent Parker bound we have two occasions in which a coherence length of turbulent magnetic fields is required. One occasion is in Eq.~\eqref{eqn:SeedField} where we estimate the seed-field amplitude of the lowest eigenmode of the mean-field dynamo by the residual average of many turbulent magnetic field cells. The length scale needed is thus the size of a turbulent magnetic correlation cell entering the volume average of the random field. Strictly speaking, this quantity is controlled by the three-dimensional correlation volume of the fluctuation field, rather than by any single one-dimensional correlation length. We therefore use $L_\mathrm{int}$ as a practical proxy which provides a physically meaningful measure of the typical spatial scale of the saturated fluctuation field. This choice is also consistent with numerical results showing that, in the saturated state of the fluctuation dynamo, $L_\mathrm{int}$ is substantially smaller than the velocity integral scale and hence better characterizes the magnetic correlation cell relevant to Eq.~\eqref{eqn:SeedField}.

The other occasion arises when we compute turbulent pre-acceleration of monopoles, for which we use an effective coherence length $l_\mathrm{eff}$. In the following we show that $l_\mathrm{eff}$ is more directly linked to $L_L$; the appropriate relation is
\begin{align}
l_\mathrm{eff}=4L_L.
\label{eqn:leff}
\end{align}

The kinetic energy gained by a monopole with magnetic charge $g$ traversing a region of turbulent magnetic fields can be written as
\begin{align}
\Delta K(L)=g\int_0^L B_t(s)\mathrm{d}s.
\end{align}
Here $B_t(s)$ denotes the component of turbulent magnetic fields projected tangentially along the monopole trajectory with $s$ being the curve length parameter, and $L$ denotes the total length of the trajectory under consideration. The mean squared value of $\Delta K(L)$ reads
\begin{align}
\left\langle [\Delta K(L)]^2 \right\rangle =
g^2 \int_0^L \mathrm{d}s \int_0^L \mathrm{d}s'\, C_t(s'-s)=2g^2 \int_0^L \mathrm{d}s \int_s^L \mathrm{d}s'\, C_t(s'-s),
\label{eqn:DKL2}
\end{align}
where we have introduced
\begin{align}
C_t(s'-s)\equiv\left\langle B_t(s)\, B_t(s') \right\rangle
\end{align}
for statistically homogeneous turbulent magnetic fields, and in the second step of Eq.~\eqref{eqn:DKL2} we have used the fact that $C_t(s'-s)=C_t(s-s')$. Introducing $\ell=s'-s$, Eq.~\eqref{eqn:DKL2} then becomes
\begin{align}
\left\langle [\Delta K(L)]^2 \right\rangle =2g^2 \int_0^L \mathrm{d}s \int_0^{L-s} C_t(\ell)\mathrm{d}\ell=2g^2 \int_0^L \mathrm{d}\ell \int_0^{L-\ell} C_t(\ell)\mathrm{d}s,
\end{align}
where in the second step the order of integration is changed by using an equivalent way to define the integration region. Since $C_t(\ell)$ is independent of $s$, the inner integration can be performed and we obtain
\begin{align}
\left\langle [\Delta K(L)]^2 \right\rangle =2g^2 \int_0^L (L-\ell)C_t(\ell)\mathrm{d}\ell.
\end{align}
In the monopole pre-acceleration problem, $L$ is much larger than the correlation length of the turbulent magnetic fields. This implies the above equation can be approximated as
\begin{align}
\left\langle [\Delta K(L)]^2 \right\rangle\simeq 2g^2 L \int_0^\infty C_t(\ell)\mathrm{d}\ell.
\end{align}
We now introduce the one-dimensional magnetic correlation length along the trajectory
\begin{align}
L_{\mathrm{tr}} \equiv \frac{1}{C_t(0)} \int_0^\infty C_t(\ell)\, d\ell
= \frac{1}{\langle B_t^2 \rangle} \int_0^\infty C_t(\ell)\, d\ell ,
\end{align}
whose definition is quite similar to $L_L$ defined in Eq.~\eqref{eqn:IntScales}. We assume $L_L$ and $L_{\mathrm{tr}}$ can be reasonably identified, since in turbulent pre-acceleration, usually the trajectory of a monopole will gain a large deflection angle when $s$ has increased by an amount much larger than the correlation length. This allows us to write
\begin{align}
\left\langle [\Delta K(L)]^2 \right\rangle \simeq 2g^2\langle B_t^2 \rangle L L_L.
\label{eqn:DKM1}
\end{align}
On the other hand, the average change in the kinetic energy of a monopole after it traverses $N_t=L/l_\mathrm{eff}$ cells of turbulent magnetic fields can be derived using the method presented in the Appendix of Ref.~\cite{Kobayashi:2023ryr}. If the kinetic energy of a monopole only changes slightly after its journey through the turbulent magnetic field cells, then the effect of turbulent pre-acceleration on the seed-field Parker bound can be neglected. Therefore, in the following when we analyze $l_\mathrm{eff}$, we will only be concerned with the situation in which the kinetic energy of a monopole changes significantly during turbulent pre-acceleration. Using results in Ref.~\cite{Kobayashi:2023ryr} we may write
\begin{align}
\Delta K_\mathrm{rms} \simeq g\sqrt{\langle B_t^2 \rangle}l_\mathrm{eff}\sqrt{\frac{N_t}{2}},
\end{align}
whose square gives (using $N_t=L/l_\mathrm{eff}$)
\begin{align}
\left\langle [\Delta K(L)]^2 \right\rangle \simeq \frac{1}{2}g^2\langle B_t^2 \rangle L l_\mathrm{eff}.
\label{eqn:DKM2}
\end{align}
Matching of Eq.~\eqref{eqn:DKM1} and Eq.~\eqref{eqn:DKM2} leads to Eq.~\eqref{eqn:leff} employed in our calculation.

\section{Appendix B. Coherence Length of the Lowest Eigenmode of the Galactic Mean-Field Dynamo}\label{sec:appB}
In the mean-field dynamo framework, the ordered galactic magnetic field is described by large-scale eigenmodes that are localized in specific radial regions, rather than by a uniform coherent patch extending over the entire disk. For the Parker-bound estimate, the relevant length is the distance over which this ordered field can coherently act on magnetic monopoles. We therefore identify the effective coherence length with the radial localization width of the fundamental growing radial eigenmode, rather than with the turbulent magnetic correlation length or the full galactic disk radius.

We follow the radial Wentzel--Kramers--Brillouin (WKB) formulation of Ref.~\cite{Ruzmaikin1985,Ruzmaikin:1988mfg}. In their thin-disk treatment one defines
\begin{align}
\lambda\equiv \frac{h_0}{r_0},
\qquad
r_* \equiv \frac{r}{r_0},
\label{eq:lambda}
\end{align}
where $h_0$ is the effective half-thickness of the ionized gas disk and $r_0$ is the radius of the magnetized region of the galactic disk.
Since $h_0\ll r_0$, the small parameter $\lambda\ll1$ allows the local
vertical dynamo eigenvalue problem to be separated from the slower radial variation of the mode. The radial eigenfunction $Q(r_*)$ then satisfies
\begin{align}
\lambda^2 \frac{\mathrm{d}}{\mathrm{d}r_*} \left[
\frac{1}{r_*}\frac{\mathrm{d}}{\mathrm{d}r_*}(r_*Q) \right]
+ [\gamma(r_*)-\Gamma_*]Q=0,
\label{eq:radial}
\end{align}
where $\gamma(r_*)$ is the local growth rate obtained from the vertical problem and $\Gamma_*$ is the global growth rate of the radial mode. This equation can be mapped to the following one-dimensional Schr\"{o}dinger equation of quantum mechanics
\begin{align}
-\frac{\hbar^2}{2M}\frac{\mathrm{d}^2\psi}{\mathrm{d}x^2}
+V(x)\psi
=E\psi,
\end{align}
which describes a quantum particle of mass $M$ and energy $E$ moving in a one-dimensional potential $V(x)$. The correspondence is given by
\begin{align}
x\leftrightarrow r_*,
\qquad
\psi(x)\leftrightarrow \sqrt{r_*}Q(r_*),
\qquad
\frac{\hbar}{\sqrt{2M}}\leftrightarrow \lambda,
\qquad
V(x)\leftrightarrow -\gamma(r_*),
\qquad
E\leftrightarrow -\Gamma_*.
\end{align} 
(A correction in the $V(x)$ correspondence relation proportional to $\lambda^2$ is neglected.) Therefore, for each radial mode $n$, the eigenvalue $\Gamma_{*n}$ can be determined from the quantization condition found in the WKB method
\begin{align}
\int_{c_n}^{d_n} [\gamma(r_*)-\Gamma_{*n}]^{1/2}\mathrm{d}r_*
=\pi\lambda\left(n-\frac12\right),
\qquad n=1,2,3,\ldots,
\label{eq:wkb}
\end{align}
with turning points fixed by
\begin{align}
\gamma(c_n)=\gamma(d_n)=\Gamma_{*n}.
\label{eq:turning}
\end{align}
Here we adopt the numbering convention $n=1,2,3,\ldots$ for the radial modes, with $n=1$ corresponding to the fundamental mode.
The WKB solution is localized mainly in the allowed region $c_n<r_*<d_n$. Hence, for the lowest radial mode,
\begin{align}
\Delta r = r_0(d_1-c_1)
\label{eq:width}
\end{align}
is the natural radial localization width. We identify this quantity with the effective coherence length of the lowest eigenmode of the galactic mean-field dynamo, which is used to set the self-consistent Parker bound.

In Ref.~\cite{Ruzmaikin1985}, the full profile $\gamma(r_*)$ is constructed from galaxy-dependent inputs such as the rotation curve, disk structure, turbulent transport coefficients, and local dynamo parameters. The eigenvalue $\Gamma_{*n}$ and the turning points are then obtained from the WKB condition. Since the Parker-bound estimate merely requires the characteristic width of the lowest radial mode, we use the same WKB framework but approximate the relevant
part of $\gamma(r_*)$ near its maximum by a local parabola,
\begin{align}
\gamma(r_*)\simeq \gamma_m-\frac{\zeta}{2}(r_*-r_m)^2,
\qquad \zeta\equiv-\gamma''(r_m)>0.
\label{eq:parabolic}
\end{align}
With
\begin{align}
x\equiv r_*-r_m, \qquad A_n\equiv \gamma_m-\Gamma_{*n},
\label{eq:xA}
\end{align}
the turning points are
\begin{align}
x_{n,\pm}=\pm\sqrt{\frac{2A_n}{\zeta}},
\qquad \Delta r_*^{(n)}
= 2\sqrt{\frac{2A_n}{\zeta}}.
\label{eq:turning-parabolic}
\end{align}
For the parabolic profile, the WKB integral becomes
\begin{align}
\int_{-\sqrt{2A_n/\zeta}}^{\sqrt{2A_n/\zeta}}
\left(A_n-\frac{\zeta x^2}{2}\right)^{1/2}\mathrm{d}x
=\frac{\pi A_n}{\sqrt{2\zeta}}.
\label{eq:integral}
\end{align}
Combining Eqs.~\eqref{eq:wkb} and \eqref{eq:integral} we may express $A_n$ in terms of $\zeta$
\begin{align}
A_n=\lambda\left(n-\frac{1}{2}\right)\sqrt{2\zeta},\qquad n=1,2,3,\ldots
\label{eq:An}
\end{align}
For the lowest radial mode, $n=1$, we obtain
\begin{align}
\Delta r = r_0 \Delta r_*^{(1)} = C_\zeta (r_0h_0)^{1/2},
\label{eq:dr}
\end{align}
where
\begin{align}
C_\zeta\equiv 2^{5/4}\zeta^{-1/4}.
\label{eq:Czeta}
\end{align}
The coefficient $C_\zeta$ encodes the local curvature of the WKB well and is an $O(1)$ factor. A rough quadratic estimate from the peaks of the $\gamma(r)$ profiles in Figs.~3 and 4 of Ref.~\cite{Ruzmaikin1985} gives $\zeta \sim 10$--$100$, corresponding to $C_\zeta \simeq 0.8$--$1.3$, thanks to the weak $\zeta^{-1/4}$ dependence that makes the result insensitive to the precise curvature value. Therefore, the uncertainty in $\zeta$ affects only the prefactor $C_\zeta$ and does not alter the leading-order scaling $\Delta r \sim \sqrt{r_0 h_0}$.

\section{Appendix C. Explicit Separation of the Turbulent and Large-Scale Fields from Averaged Recursion Relations}\label{sec:appC}

The evolution of galactic magnetism involves both the large-scale coherent fields and the small-scale turbulent fields. In this work the lowest eigenmode of the galactic mean-field dynamo at the seed stage is selected to set the self-consistent Parker bound. In this appendix we explicitly show that neglecting the impact of small-scale turbulent fields on monopole energy extraction (apart from the pre-acceleration effect which has been taken into account) is legitimate. We also comment on the validity of neglecting other eigenmodes of the large-scale galactic dynamo.

First we consider a magnetic monopole which traverses the localization region of the lowest eigenmode of the galactic mean-field dynamo. This localization region also contains a large number of turbulent cells with small-scale turbulent magnetic fields. Let the relativistic three-momentum of the monopole after passing the $n$th cell be
\begin{align}
\bm p_n = \gamma_n m \bm v_n,
\end{align}
with $\bm v_n$ and $\gamma_n$ being the three-velocity and the corresponding Lorentz boost factor of the monopole after passing the $n$th cell. The corresponding dimensionless momentum is introduced as (natural units are adopted, so that $c=1$)
\begin{align}
\bm q_n \equiv \frac{\bm p_n}{m} = \gamma_n \bm v_n.
\end{align}
In the $n$th cell, the total magnetic field $\bm B_{\mathrm{tot},n}$ can be decomposed as
\begin{align}
\bm B_{\mathrm{tot},n}=\bm B_{\mathrm L,n}+\bm b_n,
\end{align}
where $\bm B_{\mathrm L,n}$ is the large-scale eigenmode field at the monopole's current position and $\bm b_n$ is the small-scale turbulent field in the current cell.

Define the corresponding dimensionless momentum increments by
\begin{align}
\bm a_n \equiv \frac{g\tau_n}{m}\bm B_{\mathrm L,n},
\qquad
\bm\xi_n \equiv \frac{g\tau_n}{m}\bm b_n,
\end{align}
where $\tau_n$ is the time required to cross the $n$th cell. The physical system then obeys
\begin{align}
\bm q_n=\bm q_{n-1}+\bm a_n+\bm\xi_n.
\label{eq:physical-recursion}
\end{align}

Now construct a ``pure-turbulence auxiliary system.'' It has the same initial conditions as the physical system and is paired with it using the same realization of the turbulent field, but the large-scale field is artificially switched off. Denote its momentum by $\bm q_n^{(t)}$. It satisfies
\begin{align}
\bm q_n^{(t)}=\bm q_{n-1}^{(t)}+\bm\xi_n.
\label{eq:turbulent-recursion}
\end{align}
Define the additional momentum correction produced by the large-scale field:
\begin{align}
\delta\bm q_n^{(\mathrm L)}
\equiv \bm q_n-\bm q_n^{(t)}.
\label{eq:deltaq-definition}
\end{align}

We use angle brackets $\langle\cdots\rangle$ to denote a classical statistical average over the initial conditions and directions of motion of a large population of monopoles (including anti-monopoles), and also over all possible realizations of small-scale turbulent magnetic fields.

Suppose that the turbulent field in the current cell is unbiased for all given preceding trajectories, and that the pure-turbulence momentum has no systematic alignment with the local large-scale field. The following averaged cross terms then vanish:
\begin{align}
\bigl\langle \bm q_{n-1}\cdot\bm\xi_n\bigr\rangle=\bigl\langle \bm q_{n-1}^{(t)}\cdot\bm\xi_n\bigr\rangle=0,
\qquad
\bigl\langle \bm a_n\cdot\bm\xi_n\bigr\rangle=0,
\qquad
\bigl\langle \bm q_{n-1}^{(t)}\cdot\bm a_n\bigr\rangle=0.
\label{eq:zerocross}
\end{align}

Squaring Eq.~\eqref{eq:physical-recursion} and averaging (with the help of Eq.~\eqref{eq:zerocross}) gives the recursion relation for the physical system:
\begin{align}
\bigl\langle |\bm q_n|^2\bigr\rangle
-\bigl\langle |\bm q_{n-1}|^2\bigr\rangle
& =\bigl\langle |\bm\xi_n|^2\bigr\rangle
+\bigl\langle |\bm a_n|^2\bigr\rangle + 2\bigl\langle
\delta\bm q_{n-1}^{(\mathrm L)}\cdot\bm a_n
\bigr\rangle.
\label{eq:physical-mean-recursion}
\end{align}
Similarly, squaring and averaging the pure-turbulence recursion
\eqref{eq:turbulent-recursion} yields
\begin{align}
\bigl\langle |\bm q_n^{(t)}|^2\bigr\rangle
-\bigl\langle |\bm q_{n-1}^{(t)}|^2\bigr\rangle
=\bigl\langle |\bm\xi_n|^2\bigr\rangle.
\label{eq:turbulent-mean-recursion}
\end{align}
Equation~\eqref{eq:turbulent-mean-recursion} contains precisely the pure turbulent random-acceleration term that appears in
Eq.~\eqref{eq:physical-mean-recursion}.

Define
\begin{align}
D_n\equiv
\bigl\langle |\bm q_n|^2\bigr\rangle
-\bigl\langle |\bm q_n^{(t)}|^2\bigr\rangle.
\label{eq:D-definition}
\end{align}
Subtracting Eq.~\eqref{eq:turbulent-mean-recursion} from
Eq.~\eqref{eq:physical-mean-recursion} gives
\begin{align}
D_n-D_{n-1}
=\bigl\langle |\bm a_n|^2\bigr\rangle
+2\bigl\langle
\delta\bm q_{n-1}^{(\mathrm L)}\cdot\bm a_n
\bigr\rangle.
\label{eq:D-recursion}
\end{align}
The pure-turbulence term $\langle|\bm\xi_n|^2\rangle$ has canceled exactly.

On the other hand, subtracting Eq.~\eqref{eq:turbulent-recursion} from
Eq.~\eqref{eq:physical-recursion} gives
\begin{align}
\delta\bm q_n^{(\mathrm L)}
=\delta\bm q_{n-1}^{(\mathrm L)}+\bm a_n.
\label{eq:deltaq-recursion}
\end{align}
Squaring Eq.~\eqref{eq:deltaq-recursion} and averaging yields
\begin{align}
\bigl\langle |\delta\bm q_n^{(\mathrm L)}|^2\bigr\rangle
-\bigl\langle |\delta\bm q_{n-1}^{(\mathrm L)}|^2\bigr\rangle
&=
\bigl\langle |\bm a_n|^2\bigr\rangle + 2\bigl\langle
\delta\bm q_{n-1}^{(\mathrm L)}\cdot\bm a_n
\bigr\rangle.
\label{eq:deltaq-square-recursion}
\end{align}
The right-hand sides of Eqs.~\eqref{eq:D-recursion} and
\eqref{eq:deltaq-square-recursion} are identical. Therefore,
\begin{align}
D_n-D_{n-1}
= \bigl\langle |\delta\bm q_n^{(\mathrm L)}|^2\bigr\rangle
-\bigl\langle |\delta\bm q_{n-1}^{(\mathrm L)}|^2\bigr\rangle.
\label{eq:matched-recursions}
\end{align}
The physical and auxiliary systems have the same initial momentum, so
\begin{align}
\delta\bm q_0^{(\mathrm L)}=0,
\qquad D_0=0.
\end{align}
For a magnetic monopole that has traversed $N$ cells, summing Eq.~\eqref{eq:matched-recursions} from $n=1$ to $n=N$ produces telescoping sums on both sides and gives
\begin{align}
\bigl\langle |\bm q_N|^2\bigr\rangle
-\bigl\langle |\bm q_N^{(t)}|^2\bigr\rangle
=\bigl\langle |\delta\bm q_N^{(\mathrm L)}|^2\bigr\rangle.
\label{eq:separation-deltaq}
\end{align}
Iterating Eq.~\eqref{eq:deltaq-recursion} gives
\begin{align}
\delta\bm q_N^{(\mathrm L)}=\sum_{n=1}^{N}\bm a_n.
\label{eq:deltaq-sum}
\end{align}
Hence the general explicit separation formula is
\begin{align}
\bigl\langle |\bm q_N|^2\bigr\rangle
-\bigl\langle |\bm q_N^{(t)}|^2\bigr\rangle
=\left\langle
\left|\sum_{n=1}^{N}\bm a_n\right|^2
\right\rangle.
\label{eq:general-separation}
\end{align}
Thus, the increase in the mean squared momentum of the full system relative to the pure-turbulence system is exactly the mean square of the accumulated momentum correction due to the large-scale field.

In setting the self-consistent Parker bound, we assume that the large-scale field is approximately uniform throughout the
crossing region with field strength $\bm B_{\mathrm L}$ and magnitude $B$, and the monopole's crossing speed is approximately constant and equal to $v_i$. The total crossing length is $\Delta r=Nl_{\rm eff}$. This leads to
\begin{align}
\sum_{n=1}^{N}\bm a_n
=\frac{g\Delta r}{mv_i}\bm B_{\mathrm L}.
\end{align}
In this case Eq.~\eqref{eq:general-separation} becomes
\begin{align}
\bigl\langle |\bm q_N|^2\bigr\rangle
-\bigl\langle |\bm q_N^{(t)}|^2\bigr\rangle
=\left(\frac{gB\Delta r}{mv_i}\right)^2.
\label{eq:gs2}
\end{align}
In this work, we restrict our attention to non-relativistic magnetic monopoles so the monopole kinetic energy is related to
its dimensionless momentum as $E_{\mathrm{kin}}=\frac{1}{2}mq^2$. The mean additional kinetic energy produced by the large-scale eigenmode is therefore
\begin{align}
\bigl\langle\Delta E_{\mathrm L}\bigr\rangle =\frac{(gB\Delta r)^2}{2mv_i^2}.
\label{eq:energy-v}
\end{align}
Using $\gamma_i-1\simeq\frac{v_i^2}{2}$, this can also be written as
\begin{align}
\bigl\langle\Delta E_{\mathrm L}\bigr\rangle
=\frac{(gB\Delta r)^2}{4m(\gamma_i-1)}.
\label{eq:energy-gamma}
\end{align}
This is exactly the energy extraction formula employed to obtain the third term on the right-hand side of Eq.~\eqref{eqn:Bevolve} and the self-consistent Parker bound.

In the preceding derivation, we retained only the lowest eigenmode of the galactic mean-field dynamo. Unlike the contribution of the small-scale turbulent field, the monopole-induced cross terms between distinct large-scale eigenmodes do not generally vanish under ensemble averaging. Neglecting higher modes is nevertheless justified when their localization regions are well separated from that of the fundamental mode, or when their growth rates are sufficiently smaller that the fundamental mode rapidly dominates. The most challenging situation arises when a second mode has a comparable growth rate and a substantially overlapping localization region. A rigorous treatment would then require coupled evolution equations for the two mode amplitudes, including the off-diagonal monopole-response terms. Since such nearly degenerate modes have similar spatial and temporal scales, we do not expect them to modify the resulting Parker bound by orders of magnitude. A quantitative multimode analysis is left for future work.

\section{Appendix D. Geometric Factor of an Annular Cylindrical Shell}\label{sec:appD}

Let $D$ be the localization region of the lowest mean-field dynamo eigenmode, with boundary $\partial D$, volume $V$, and total boundary area $A$. In the one-zone description, $B(t)$ denotes the characteristic strength of this eigenmode throughout $D$. Its magnetic energy $E_B$ and the corresponding time derivative $\dot E_B$ can be expressed as (in Gaussian units)
\begin{align}
E_B=\frac{B^2}{8\pi}V,
\qquad
\dot E_B=\frac{VB}{4\pi}\frac{\dd B}{\dd t}.
\label{eq:magnetic_energy}
\end{align}
Let $F_s$ be the local differential monopole flux at the seed stage, and assume that it is approximately homogeneous and isotropic near $\partial D$.  For a boundary element $\dd A$ with outward unit normal $\hat{\bm n}$, the rate of inward crossings is
\begin{align}
\dd\dot N_{\rm in} = F_s\,\dd A \int_{\hat{\bm v}\cdot\hat{\bm n}<0}
\left|\hat{\bm v}\cdot\hat{\bm n}\right|\dd\Omega = \pi F_s\,\dd A,
\label{eq:local_crossing_rate}
\end{align}
because
\begin{align}
\int_{\hat{\bm v}\cdot\hat{\bm n}<0}
\left|\hat{\bm v}\cdot\hat{\bm n}\right|\dd\Omega
=2\pi\int_0^{\pi/2}\cos\theta\sin\theta\,\dd\theta
=\pi.
\end{align}
Consequently,
\begin{align}
\dot N_{\rm in}=\pi A F_s.
\label{eq:total_crossing_rate}
\end{align}
No spherical or convex geometry was used.  For a multiply connected or nonconvex region, Eq.~\eqref{eq:total_crossing_rate} counts inward boundary crossings, or equivalently the beginnings of magnetized traversal segments.

More generally, if a crossing at $\bm x\in\partial D$ with incident direction $\hat{\bm v}$ extracts energy
$\Delta E(\bm x,\hat{\bm v})$, then the monopole contribution to $\dot E_B$ is given by
\begin{align}
\left.\dot E_B\right|_M =
-F_s\int_{\partial D}\dd A
\int_{\hat{\bm v}\cdot\hat{\bm n}<0}
\left|\hat{\bm v}\cdot\hat{\bm n}\right|
\Delta E(\bm x,\hat{\bm v})\,\dd\Omega.
\label{eq:general_loss}
\end{align}
We will be mostly interested in the case in which $\Delta E(\bm x,\hat{\bm v})$ is approximately independent of $\bm x$ and $\hat{\bm v}$. In such a case, the above equation simplifies to
\begin{align}
\left.\dot E_B\right|_M =-\pi A F_s\Delta E.
\label{eq:constant_loss}
\end{align}

Define
\begin{align}
\mu\equiv\frac{2m(\gamma_i-1)}{g\Delta r}.
\label{eq:mu}
\end{align}
The interpolation
\begin{align}
	\Delta E(B)\simeq
	\frac{gB\Delta r}{1+2\mu/B}
	\label{eq:energy_interpolation}
\end{align}
connects the slight- and significant-deviation limits:
\begin{align}
\Delta E\simeq
	\begin{cases}
		\displaystyle
		\frac{(gB\Delta r)^2}{4m(\gamma_i-1)},
		& B\ll 2\mu,\\[1.0ex]
		gB\Delta r,
		& B\gg 2\mu.
	\end{cases}
\label{eq:energy_limits}
\end{align}

The one-zone approximation replaces the direction- and position-dependent energy gain in
Eq.~\eqref{eq:general_loss} by Eq.~\eqref{eq:energy_interpolation} for every
inward crossing. Eqs.~\eqref{eq:constant_loss} and \eqref{eq:energy_interpolation} then give
\begin{align}
\left.\dot E_B\right|_M
= -\pi A F_s\,
\frac{gB\Delta r}{1+2\mu/B}.
\label{eq:loss_one_zone}
\end{align}
Combining Eqs.~\eqref{eq:magnetic_energy} and
\eqref{eq:loss_one_zone} yields
\begin{align}
\left.\frac{\dd B}{\dd t}\right|_M
= -4\pi^2\frac{A\Delta r}{V}
\frac{F_sg}{1+2\mu/B}.
\label{eq:B_loss_raw}
\end{align}
It is convenient to define
\begin{align}
Q\equiv\frac{3V}{A\Delta r},
\label{eq:Q_general}
\end{align}
so that
\begin{align}
\left.\frac{\dd B}{\dd t}\right|_M
=
-\frac{12\pi^2}{Q}
\frac{F_sg}{1+2\mu/B}.
\label{eq:B_loss_Q}
\end{align}
The factor $3$ normalizes the result to $Q=1$ for a sphere of radius
$\Delta r$, for which $V=(4\pi/3)\Delta r^3$ and $A=4\pi\Delta r^2$.

For the lowest eigenmode of the galactic mean-field dynamo, we approximate the localization region as an annular cylindrical shell
\begin{align}
r_-\leq r\leq r_+,
\qquad
-h_0\leq z\leq h_0,
\qquad
r_\pm\equiv r_s\pm\frac{\Delta r}{2},
\label{eq:shell_definition}
\end{align}
where $r_s$ is the localization radius, $\Delta r=r_+-r_-$ is the radial
width, and $2h_0$ is the full disk thickness. Its volume is
\begin{align}
V = 2h_0\pi\left(r_+^2-r_-^2\right) = 4\pi r_sh_0\Delta r.
\label{eq:shell_volume}
\end{align}
The combined area of the inner and outer cylindrical faces is
\begin{align}
A_{\rm side}
=2\pi r_+(2h_0)+2\pi r_-(2h_0)
=8\pi r_sh_0,
\label{eq:side_area}
\end{align}
whereas the two annular end faces have total area
\begin{align}
A_{\rm end}
=2\pi\left(r_+^2-r_-^2\right)
=4\pi r_s\Delta r.
\label{eq:end_area}
\end{align}
Hence
\begin{align}
A=4\pi r_s(\Delta r+2h_0).
\label{eq:shell_area}
\end{align}
Substitution into Eq.~\eqref{eq:Q_general} gives
\begin{align}
Q = \frac{3V}{A\Delta r} = \frac{3h_0}{\Delta r+2h_0}.
\label{eq:Q_shell}
\end{align}
The cancellation of $r_s$ is exact for the ideal sharp-edged shell defined in Eq.~\eqref{eq:shell_definition}. The result is approximate only insofar as the smooth eigenmode is replaced by this one-zone shell and all traversal segments are assigned the same effective interaction length $\Delta r$. For the fiducial values $h_0=0.5\,\mathrm{kpc}$ and $\Delta r\simeq\sqrt{r_0h_0}$ with $r_0=9\,\mathrm{kpc}$, one finds
$Q\simeq0.48$.

\end{document}